\documentclass[aps,prb,reprint,groupedaddress]{revtex4-1}
\usepackage[latin1]{inputenc}
\usepackage{amsmath}
\usepackage{amsfonts}
\usepackage{amssymb}
\usepackage{graphicx}
\usepackage{geometry}
\begin{document}
\title{Pressure study of nematicity and quantum criticality in Sr$_3$Ru$_2$O$_7$ for an in-plane field}
\author{Dan Sun$^1$, W. Wu$^1$, S. A. Grigera$^2$, R. S. Perry$^3$, A. P. Mackenzie$^{2,4,5}$, and S. R. Julian$^{1,4}$}

\affiliation{$^1$ Department of Physics, University of Toronto, 60 St.\ George Street, Toronto, M5S 1A7, Canada\\
$^2$ Scottish Universities Physics Alliance, School of Physics and Astronomy, University of St.\ Andrews, North Haugh, St.\ Andrews KY16 9SS, United Kingdom\\
$^3$ Center for Science at Extreme Conditions, School of Physics, University of Edinburgh, Edinburgh, EH9 3JZ, Scotland\\
$^4$ Canadian Institute for Advanced Research, 180 Dundas St.\ W., Suite 1400, Toronto, M5G 1Z8, Canada\\
$^5$ Max-Planck Institute for Chemical Physics of Solids, Noethnitzerstr.\ 40,  Dresden, 01187, Germany} 
\date{\today}
\begin{abstract}
We study the relationship between the nematic phases of Sr$_3$Ru$_2$O$_7$ and quantum criticality. At ambient pressure,  one nematic phase is associated with a metamagnetic quantum critical end point (QCEP) when the applied magnetic field is near the \textit{c}-axis. We show, however, that this metamagnetic transition does not produce the same nematic signatures when the QCEP is reached by hydrostatic pressure with the field applied in the \textit{ab}-plane. Moreover, a second nematic phase, that is seen for field applied in the \textit{ab}-plane close to, but not right at, a second metamagnetic anomaly, persists with minimal change to the highest applied pressure, 16.55 kbar. 
Taken together our results suggest that metamagnetic quantum criticality 
  may not be necessary for the formation of a nematic phase in Sr$_3$Ru$_2$O$_7$. 
\end{abstract}
\maketitle

\section{Introduction}
Electronic nematic phases are electronic analogues of nematic liquid crystals. They are characterized by spontaneously broken rotational symmetry in their transport properties. Such nematic phases, which have a symmetry level between that of a Fermi liquid and a Wigner solid, \cite{Fradkin2010} have been unveiled in several strongly correlated electron systems, including quantum Hall systems,\cite{Lilly1999} and iron-pnictide \cite{Chu2010a} and cuprate superconductors.\cite{Ando2002,Hinkov2008} A prominent example is Sr$_3$Ru$_2$O$_7$,\cite{Grigera2004} which is a clean system for nematicity in the sense that accompanying the nematic phase transition, there is no magnetic ordering or charge density wave formation, although an area-preserving symmetry-breaking lattice distortion of order $10^{-6}$ occurs within the nematic phase.\cite{Stingl2011} 

Sr$_3$Ru$_2$O$_7$ is the bilayer member of the Ruddlesden-Popper series of layered perovskite ruthenates. It has orthorhombic Bbcb symmetry, arising from ordered rotations of the RuO$_6$ octahedra, but is nearly tetragonal in terms of the lattice parameter: the 5 parts in 10$^4$ difference between the \textit{a} and \textit{b} lattice parameters cannot be detected by Laue X-ray diffraction.\cite{Shaked2000} The resistivity is isotropic in the two in-plane principal axis directions in the absence  of an external field.

\begin{figure}[htb]
\includegraphics[width=7.4cm]{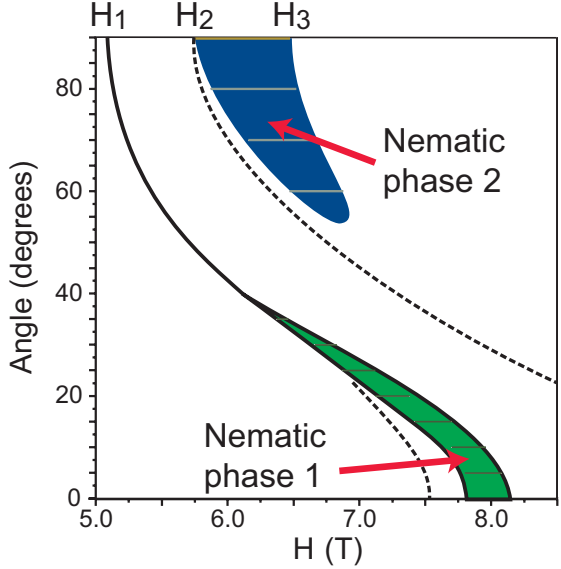} 
\caption{Phase diagram at ambient pressure 
 in the (H, $\theta$) plane at T$\ll$1 K, from Refs.\ \onlinecite{Borzi2007,Raghu2009,Mackenzie2012}. 0$^\circ$ is the $c$-axis, while 90$^\circ$ is the $ab$-plane.  The two metamagnetic transitions follow the $H_1$ and $H_2$ lines. Where the lines are solid the metamagnetic transitions are first-order, while dotted lines show crossovers. The green and blue shaded regions show nematic phase 1 and 2 respectively. 
Nematic phase 1 
 is bounded by first order transitions. 
For the $H_2$ metamagnetic transition, and nematic phase 2, 
first order behaviour has only been 
  observed at $H_2$ at the \textit{ab}-plane. $H_3$ indicates the end of nematic phase 2. 
} 
\label{fig:intro}
\end{figure}

There are two distinct nematic regions that have been found in the magnetic field vs.\ field angle phase diagram of Sr$_3$Ru$_2$O$_7$, as shown in Fig.\  \ref{fig:intro}.\footnote{Strictly speaking, whenever the field angle is not 0$^\circ$, these phases are `meta-nematic'\cite{Puetter2007} since the applied field already breaks the rotational symmetry of the crystal, but we use the term `nematic' throughout this paper.} These nematic phases can only be seen in very pure samples, at temperatures below 1 K. The  nematic phase 1 
extends between 0$^\circ$ and 40$^\circ$ from the \textit{c}-axis and is bounded by first order transitions, which are demonstrated by peaks in the imaginary part of the \textit{ac} susceptibility $\chi''(H)$.\cite{Grigera2004, Borzi2007, Rost2009}  Nematic phase 2 
 is found from 60$^\circ$ to 90$^\circ$ from  the \textit{c}-axis, i.e.\ adjacent to the $ab$-plane. 
It is bounded on each side by double peaks in the real part of the susceptibility $\chi'(H)$, which was obtained by taking the derivative, $dM/dB$, of the dc magnetization.\cite{Perry2005}
$\chi''(H)$ shows a small peak at the lower-field boundary of nematic phase 2 
 at ambient pressure at the \textit{ab}-plane.\cite{Wu2011} 
There is no observable feature in  $\chi''(H)$ at the upper-field boundary of the phase. 

\begin{figure}[htb]
\includegraphics[width=7.4cm]{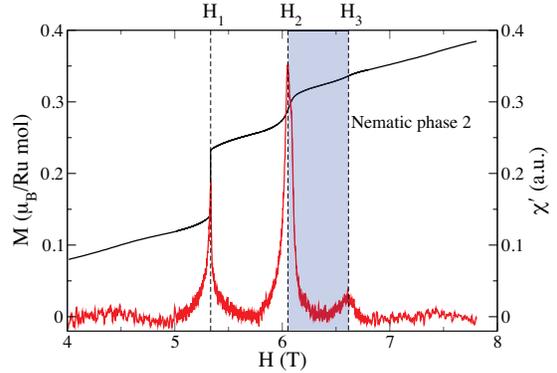} 
\caption{Low temperature susceptibility $\chi(H)$\cite{Wu2011} (red) and integrated susceptibility $M(H)$ (black) at near-ambient pressure. The two metamagnetic transitions $H_1$ and $H_2$ are indicated by dashed lines. A first order jump in $M(H)$ at $H_1$ is added by hand (see Appendix A). The weak $H_3$ anomaly appears only below $\sim$ 0.4 K.
} 
\label{fig:chi}
\end{figure}

The two nematic phases are closely related to metamagnetic features that survive to much higher temperature and that are not sensitive to sample purity. 
Metamagnetism is defined as a sudden increase in magnetization 
  within a narrow range of field $H$.
This is illustrated for Sr$_3$Ru$_2$O$_7$ in Fig.\ \ref{fig:chi},
  which shows low excitation frequency ($\nu= 14 Hz$) susceptibility data $\chi(H)$ (red line) taken at 70 mK 
  with the external field between 4 and 8 T applied in the $ab$-plane,
  together with the magnetization $M(H)$ obtained by integrating $\chi(H)$, 
  as described in Appendix A. 
The weak $H_3$ metamagnetic transition marks the upper boundary of nematic phase 2 
and, 
  in contrast to the $H_1$ and $H_2$ metamagnetic transitions, it is not robust, disappearing rapidly 
  with increasing $T$. 
The dependence of the $H_1$ and $H_2$ metamagnetic features on field-angle 
  is shown in Fig.\ \ref{fig:intro}.

The $H_1$ metamagnetic transition exhibits an interesting kind of quantum criticality. This transition is a first-order jump at low temperature for magnetic field in the \textit{ab}-plane, 
  as illustrated in Fig.\ \ref{fig:chi} at $H_1$. 
When the field is rotated towards the \textit{c}-axis, the jump decreases in magnitude, vanishing at a quantum critical end point (QCEP) approximately 10$^\circ$ from the \textit{c}-axis in samples where purity is not high enough to see the nematic phase.\cite{Grigera2003}

The fact that nematic phase 1, 
in very high purity samples, emerges near a 
  QCEP has caused speculation of a deep connection. 
A connection at some level is plausible, because 
  the first-order metamagnetic jump and nematicity are mutually exclusive ways to 
  avoid a van Hove singularity at the Fermi energy $\varepsilon_F$.\cite{Grigera2004,Kee2005} 
Puetter et al.,\cite{Puetter2010} using a tight-binding model, identified two van Hove singularities 
  close to $\varepsilon_F$. 
As a sweeping magnetic field progressively spin-splits the energy bands, the two van Hove 
  singularities will consecutively pass through $\varepsilon_F$, producing peaks in the 
  density of states, $g(\varepsilon_F)$. 
In mean-field theories for itinerant systems, the tendency to order increases as 
  $Ig(\varepsilon_F)$, where $I$ is an electron-electron interaction strength.  
In the case of Sr$_3$Ru$_2$O$_7$, 
  an intra-orbital Hubbard $U$ interaction will drive a metamagnetic transition \cite{Wohlfarth1962}, 
  while on-site inter-orbital electron repulsion \cite{Raghu2009,lee2009theory} 
  and nearest-neighbour repulsion 
  \cite{Puetter2010,Puetter2012} drive different forms of nematic order.
The key difference between the first-order metamagnetic transition and nematicity is that, in 
  the former all 
  four symmetrically placed copies of a van Hove singularity in the Brillouin zone are jumped 
  over at once, while in the latter 
  they are jumped over two-at-a-time. 
In the nematic phase, when only two of the four singularities 
  have been jumped over, the Fermi surface is strongly distorted. The point is that, because first-order metamagnetism and nematicity are mutually exclusive, weakening the strong first-order metamagnetic jump as it approaches its QCEP would naturally be a precondition for the appearance of the nematic phase. 

Beyond this mutual connection to an underlying van Hove singularity 
  the  nematic phase 1 
  gives the appearance of screening the metamagnetic QCEP, 
  just as quantum critical superconductivity is often found to screen antiferromagnetic 
  quantum critical points.\cite{mathur1998} 

We tried to investigate this connection in a previous high-pressure study, 
  by measuring the susceptibility. 
We succeeded in inducing the QCEP for the $H_1$ metamagnetic transition with 
  H$\parallel$\textit{ab}-plane, 
  at 13.6$\pm$0.2 kbar, but we did not see the bifurcation of the peak in $\chi'(H)$ that 
  is associated with the appearance of the nematic phase when $H\parallel c$.\cite{Wu2011}
This is in contrast to the theoretical prediction of Ref.\ \onlinecite{Raghu2009} that the 
  nematic phase 1 
  should extend to the $ab$-plane when it is not pre-empted by a first-order 
  metamagnetic transition. 

In contrast to nematic phase 1, 
 nematic phase 2 
 has not been considered theoretically. 
It has very similar transport signatures to nematic phase 1, 
 and 
  it is similarly sensitive to impurities and disappears above 0.4 K.\cite{Perry2005}
It is not obvious however that it screens a metamagnetic QCEP. 
The metamagnetic feature at $H_2$ is weak, moreover the nematic phase occurs not at, 
  but beside the metamagnetic transition field $H_2$, and it extends to fields that are quite a 
  lot higher (see Fig.\  \ref{fig:chi}). 
However, in heat capacity measurements at widely spaced fields, 
  $C/T$ increased logarithmically with decreasing temperature near $H_2$, 
  suggesting quantum critical behaviour.  
But it is not necessarily metamagnetic quantum criticality: 
  recent theoretical work \cite{ponte2013fermi} suggests that a nematic phase would have a 
  non-Fermi liquid normal state regardless of proximity to a metamagnetic QCEP. 

The motivation for the present study was to examine more deeply the connection between 
  nematic phases and quantum criticality. 
The plan was to use resistivity anisotropy to search again for nematicity at the $H_1$ QCEP and, 
  secondly, to see the effect of pressure on nematic phase 2. 
Pressure should reduce the peak in $g(\varepsilon_F)$ at the van Hove singularity, weakening metamagnetism, so if there is a connection to quantum critical metamagnetism 
  nematic phase 2  
  should weaken with increasing pressure. 

Our main results are that we find no evidence of a nematic phase at the QCEP of the $H_1$ 
  metamagnetic transition, while nematic phase 2  
  is robust against pressure up to 16.55 kbar.

\section{Experiment}
We simultaneously measured the resistivity of two samples, $\rho_{\parallel}$ with current parallel to the field and $\rho_{\perp}$ with current perpendicular to the field,  in the same clamp-type pressure cell. Daphne oil 7373 was used as the pressure medium and the pressure at low temperature was determined by the calibrated pressure dependence of the superconducting transition temperature of tin. The field and the current were both applied in the \textit{ab}-plane. The current directions for $\rho_{\parallel}$ and $\rho_{\perp}$ were 8.5$^\circ$ and 15.5$^\circ$ from the closest in-plane principal axis, respectively(the \textit{a} and \textit{b} axes are indistinguishable under Laue X-ray diffraction). At eight different  pressures ranging from 1.87 kbar to 16.55 kbar, we carried out field sweeps crossing both of the metamagnetic anomalies, at temperatures from 100 mK to 2.5 K.  At each of these pressures we also carried out twenty or so temperature sweeps at fixed fields, to extract the A coefficient in $\rho=\rho_0+AT^2$. These temperature sweeps went from 100 mK to 700 mK. The samples were cut from ultra pure single crystals ($\rho_{0}$ $<$ 0.4 $\mu\Omega$ cm) grown at St. Andrews University, UK. 

\begin{figure}[htb]
\includegraphics[width=7.2cm]{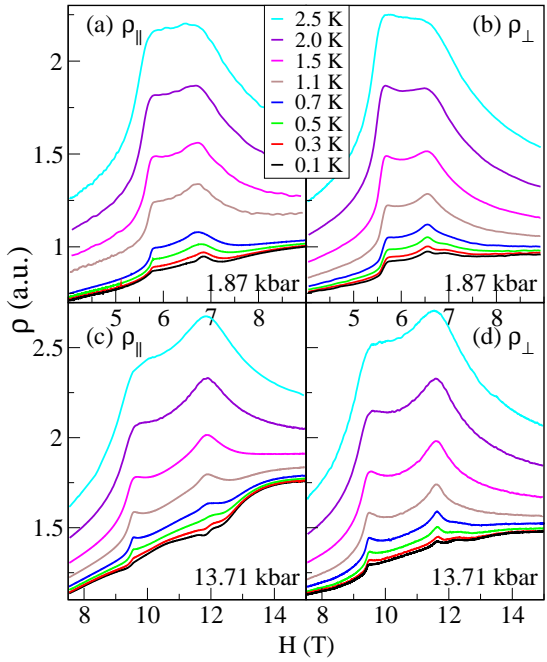} 
\caption{In-plane magnetoresistivity $\rho(H)$ with currents parallel $\rho_{\parallel}$ (left) and perpendicular $\rho_{\perp}$ (right) to the applied field at different temperatures. The upper two Figures (a) and (b) show data at 1.87 kbar and the lower two (c) and (d) show data at 13.71 kbar. $\rho_{\parallel}$ and  $\rho_{\perp}$ are normalised according to the geometry of the samples and the amplification gain. }
\label{fig:rhovsH}
\end{figure}

\section{Results}
\begin{figure}[htb]
\includegraphics[width=7.2cm]{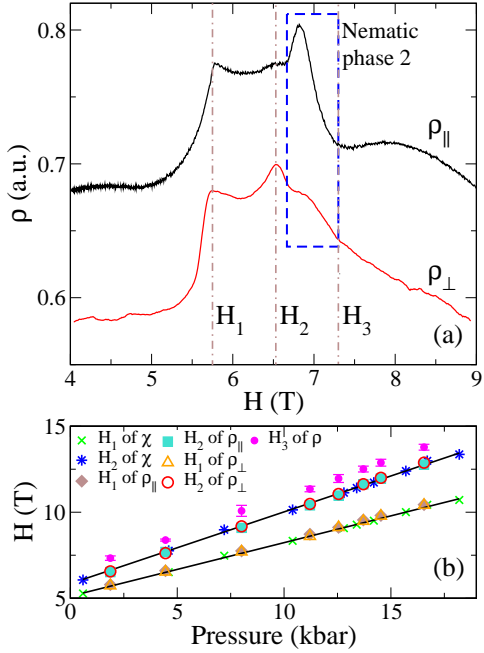} 
\caption{(a) shows $\rho_{\parallel}$ and $\rho_{\perp}$ at 100 mK  at 1.87 kbar with a linear background subtracted. The curves are normalised according to the geometry of the samples and the circuit gain and then shifted for clarity. The positions of $H_1$, $H_2$, $H_3$ and nematic phase 2  
are indicated in the figure. (b) shows the pressure dependence of the $H_1$ and $H_2$ metamagnetic anomalies and the upper boundary of nematic phase 2, 
 $H_3$. The green crosses and the blue stars show the results from earlier susceptibility measurements, and the two lines $H_1(p)$ and $H_2(p)$ are linear fits to these data. The other five sets of data are from the present resistivity measurements, as shown in the legend. The $H_3$ points were obtained by averaging the estimated upper phase boundaries from $\rho_{\parallel}$ and $\rho_{\perp}$, with error bars showing the uncertainty. }
\label{fig:1kbar}
\end{figure}

Typical in-plane magnetoresistivity data at different temperatures and pressures are shown in Fig.\  \ref{fig:rhovsH}, while Fig.\  \ref{fig:1kbar} focuses on the lowest temperature data at 1.87 kbar. 

The two metamagnetic anomalies are clearly visible. At high temperatures, these anomalies are two overlapping peaks. At low temperature (see Fig.\  \ref{fig:1kbar}(a)), the $H_{1}$ metamagnetic transition is a cusp-like feature in both $\rho_{\parallel}$ and $\rho_{\perp}$, while the $H_2$ transition is a clear peak in $\rho_{\perp}$, but a weak shoulder in $\rho_{\parallel}$.

\begin{figure}[htb]
\includegraphics[width=7.2cm]{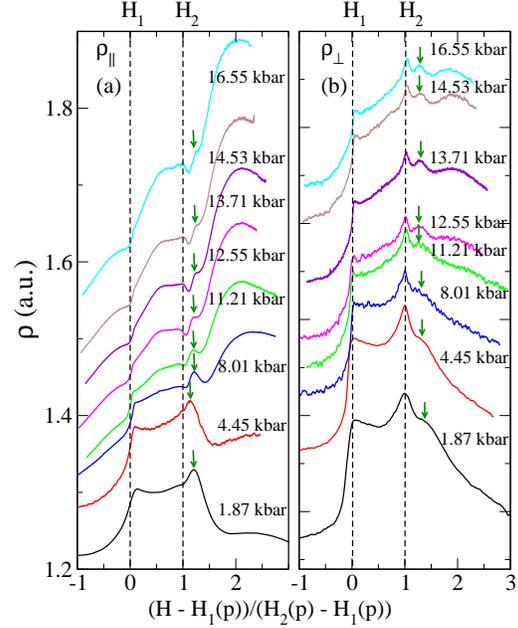} 
\caption{(a) and (b) show the magnetoresistivity, $\rho_{\parallel}$ and $\rho_{\perp}$, at 100 mK at all pressures. Figures are scaled in the field as $H\rightarrow(H-H_{1}(p))/(H_{2}(p)-H_{1}(p))$ so that the two metamagnetic transitions align, where $H_{1}(p)$ and $H_{2}(p)$ are the pressure dependence of the metamagnetic transitions extracted by fitting earlier susceptibility measurements.\cite{Wu2011}  The two dashed lines are guides to the eye for $H_1$ and $H_2$. A linear background has been subtracted at 1.87 kbar so that the resistivity returns to its original values outside the region of interest and the same background is also subtracted from the data of all the other pressures.  The curves have been  shifted vertically  for clarity. The green arrows indicate the position of nematic phase 2 at each of the pressures.}
\label{fig:100mK}
\end{figure}

\begin{figure}[htb]
\includegraphics[trim = 0mm 0mm 0mm 5mm,width=7.2cm]{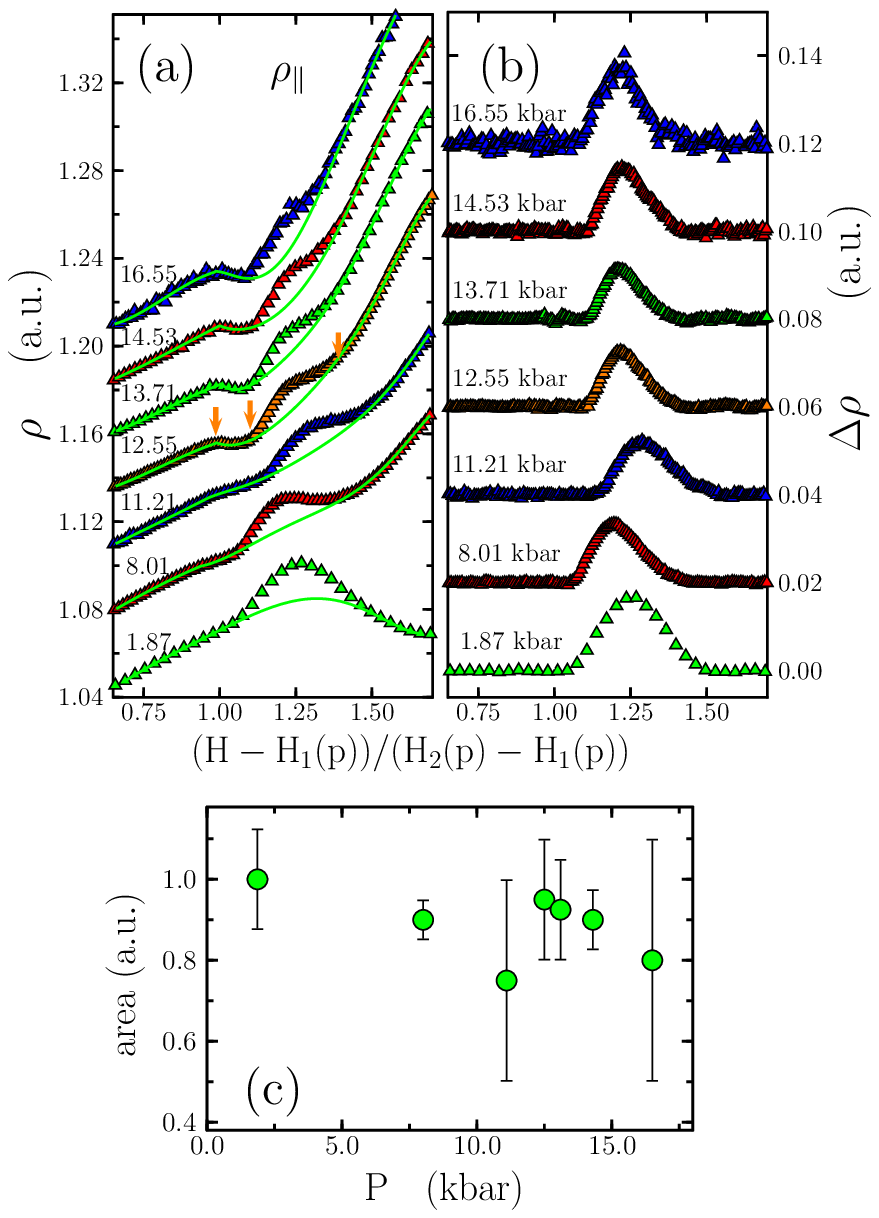} 
\caption{Pressure dependence of nematic phase 2. 
 (a) The magnetoresistivity near $H_2$. In order to estimate the pressure dependence of the nematic bump, we interpolated a background across the nematic phase as follows: The magnetoresistance was fitted by polynomials in two regions, above and below $H_2$ (indicated by the lowest arrow in the 12.5 kbar data). In fitting the region above $H_2$, the data in the nematic region were ignored (the region between the two upper arrows in the 12.5 kbar data). The data on panel (b) were obtained by subtracting the resulting fitted lines from the measured data. This procedure was adopted purely to remove a background signal, and we do not assert that the parameters of the polynomials have any particular physical significance. 
(c) This figure shows the integrated area of the resulting nematic bump. The error bars were estimated by expanding and contracting the ignored region in the fit.}
\label{fig:2ndMMT}
\end{figure}

As the pressure increases, the $H_1$ and $H_2$ transitions shift to higher fields, roughly linearly with increasing pressure, as shown in Figure \ref{fig:1kbar}(b). Fig.\  \ref{fig:1kbar}(b) also shows $H_1$ (crosses) and $H_2$ (stars) from our previous susceptibility measurements.\cite{Wu2011} By fitting the susceptibility points, we obtained linear functions $H_1(p)$ and $H_2(p)$. The features from resistivity and susceptibility align well, although the agreement is not perfect. In Fig.\  \ref{fig:100mK}, for the lowest temperature data at each pressure, we have used these $H_1(p)$ and $H_2(p)$ functions to rescale the horizontal axis using $H\rightarrow(H-H_{1}(p))/(H_{2}(p)-H_{1}(p))$. If $H_1$ and $H_2$ from our resistivity data agreed perfectly with the fit of our susceptibility results then $H_1$ would be at 0 and $H_2$ at 1 for every curve.   Some scatter is however apparent, particularly at the lowest pressures; for example in $\rho_{\parallel}$ at 1.87 kbar the $H_1$ peak is clearly above 0.

We now turn to the pressure dependence of $\rho_{\parallel}$ and $\rho_{\perp}$. At a given pressure, $\rho_{\parallel}$ and $\rho_{\perp}$  behave similarly at high temperatures. At low temperatures, however, they show qualitative differences that grow with increasing pressure. For example, as noted above, the behavior at $H_2$ is different (see Fig.\  \ref{fig:1kbar}(a)). Moreover, while $\rho_{\perp}$ only shows mild changes in shape with increasing pressure,
the low temperature curves for $\rho_{\parallel}$ change from concave upwards to concave downwards in the two regions: $H<H_1$ and $H_{1}<H<H_{2}$. For $H>H_{2}$, $\rho_{\parallel}$ is concave downwards at both low and high pressures, but the curvature is more pronounced at high pressures (see Fig.\  \ref{fig:100mK}(a)).

The small bump located just above $H_2$ corresponds to nematic phase 2, 
 discussed in the introduction \cite{Perry2005,Borzi2007} (see Fig.\  \ref{fig:1kbar}(a)). The bumps can be seen in both $\rho_{\parallel}$ and $\rho_{\perp}$, but the signal in $\rho_{\parallel}$ is stronger. The bump can only be seen at temperatures lower than $\sim$0.5 K, but it does not seem to show a strong dependence on pressure (see Fig.\  \ref{fig:2ndMMT}).

\begin{figure}[htb]
\includegraphics[width=7.2cm]{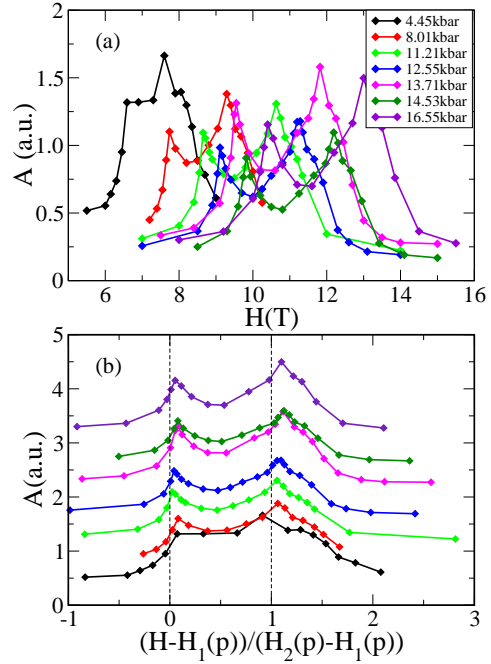} 
\caption{The $T^2$ coefficient of resistivity. Resistivity vs.\ temperature data for our $\rho_{\parallel}$ sample between 100 mK and 500 mK, were fitted with a form $\rho(T)=\rho_{\circ}+AT^2$. (a) A vs H at each pressure.  (b) A vs. H replotted with the field-axis rescaled as in Fig.\  \ref{fig:100mK} and \ref{fig:2ndMMT}, and with each plot shifted vertically for clarity.} 
\label{fig:AvsH}
\end{figure}

The \textit{A} coefficient (see Fig.\  \ref{fig:AvsH}), which is proportional to the square of the effective mass of the quasiparticles, is obtained by fitting the temperature-dependent magnetoresistivity. At $H_1$, the \textit{A} coefficient is enhanced, but not divergent even near the the critical pressure. The peak at $H_2$ and a shoulder-like feature above $H_2$ are evident. 

\section{Discussion}

The pressure dependent metamagnetic transition fields $H_1(p)$ and $H_2(p)$ of $\rho_{\parallel}$ and $\rho_{\perp}$ line up well with each other and with our susceptibility measurement,\cite{Wu2011} demonstrating hydrostatic pressure inside the cell. This is important because, although hydrostatic pressure tunes Sr$_3$Ru$_2$O$_7$ away from ferromagnetism, uni-axial stress components can induce ferromagnetism.\cite{Yaguchi2006} The extrapolation to zero pressure also agrees well with measurements at ambient pressure for $H\parallel$\textit{ab}, showing that both samples were well aligned with the field in the \textit{ab}-plane.  Although this does not ensure that the {\em current} was properly aligned parallel or perpendicular to the field, the geometry of the pressure cell constrains the current to be nearly parallel to the field for the $\rho_{\parallel}$ sample. We cannot rule out that the $\rho_{\perp}$ signal may have a small component of current parallel to the field. 

Our motivation in measuring $\rho_{\parallel}$ and $\rho_{\perp}$ was to search for signs of nematicity, however this turned out to be more complicated than we expected. 
In their study of nematic phase 1 
  with the field applied {\em near} the \textit{c}-axis, 
  Borzi et al.\ \cite{Borzi2007} defined a nematic order parameter as the anisotropy  
  $(\rho_{\parallel}-\rho_{\perp})/(\rho_{\parallel}+\rho_{\perp})$, 
  where $\rho_{\parallel}$ and $\rho_{\perp}$ referred to whether the current was 
  parallel or perpendicular to the small in-plane component of the magnetic field. 
In their configuration, because the $\sim$ 8 T field was near the \textit{c}-axis, both $\rho_{\parallel}$ and $\rho_{\perp}$ were predominantly measuring the transverse magnetoresistance and, outside of the nematic phase, they agreed well with each other. In our measurement, there are no field ranges where $\rho_{\parallel}$ and $\rho_{\perp}$ agree well, and the disagreement grows as the pressure increases. (Compare for example the lowest temperature curves in Fig.\  \ref{fig:rhovsH}(c) and (d).) That is, $(\rho_{\parallel}-\rho_{\perp})/(\rho_{\parallel}+\rho_{\perp})$ is non-zero everywhere.  This may be caused by the fact that our measurement configuration differs in two significant ways from that of Borzi et al..\ Firstly, they measured $\rho_{\parallel}$ and $\rho_{\perp}$ in the same sample. We are unable to tilt our pressure cell by 90$^\circ$, so we measured $\rho_{\parallel}$ and $\rho_{\perp}$ in different samples. We attempted to minimize the effect of this by using samples from the same batch, in the same pressure cell, but we cannot rule out a sample dependent effect, e.g., the currents are at different angles to the principal axis of the samples. Secondly, in our experiment the magnetic field is purely in the \textit{ab}-plane, so $\rho_{\parallel}$ and $\rho_{\perp}$ measure the longitudinal and transverse magnetoresistance, respectively. 
This too could contribute to  the different shape of $\rho_{\parallel}$ and $\rho_{\perp}$ at low temperature. Beyond the conventional magnetoresistivity, the applied magnetic field breaks the symmetry in the plane and could also induce `meta-nematic' anisotropy. Whatever the explanation, non-zero $(\rho_{\parallel}-\rho_{\perp})/(\rho_{\parallel}+\rho_{\perp})$ is probably not, in our measurement configuration, a reliable signature of nematicity. 
Nevertheless, a reliable signature of a nematic phase could be an abrupt {\em increase} in 
  $(\rho_{\parallel}-\rho_{\perp})/(\rho_{\parallel}+\rho_{\perp})$ 
  on entering a nematic phase,\cite{Puetter2007, Raghu2009}  
  as we see at the boundaries of nematic phase 2.
 It is therefore significant that no dramatic change in anisotropy is seen at $H_1$, even near the critical pressure, 13.6 kbar of the QCEP (see Fig.\  \ref{fig:100mK}(a) and (b)). Thus our magnetoresistance measurement provides no evidence of a nematic phase at the pressure-induced QCEP of $H_1$.

The bump in the resistivity just above $H_2$ was shown by Borzi et al.\ \cite{Borzi2007} to correspond to a nematic phase with a strong $(\rho_{\parallel}-\rho_{\perp})/(\rho_{\parallel}+\rho_{\perp})$ signature at ambient pressure. We find that this bump is robust against pressure, as seen in Figs.\  \ref{fig:100mK} and \ref{fig:2ndMMT}: surprisingly, after subtraction of the background by interpolating across the bump, the size of the peak does not change with pressure (Fig.\  \ref{fig:2ndMMT}(b) and (c)), within the error. 

In our earlier susceptibility study,\cite{Wu2011} we found the peaks in the real part of $\chi(H)$ that mark the boundary of this phase. The peak at the lower boundary had the double feature  observed by Perry et al.\ \cite{Perry2005} (see Fig.\  \ref{fig:Appchi}) and its amplitude depended only weakly on pressure,\cite{Wu2011} which is consistent with this nematic phase being relatively unaffected by pressure.  

Nematic phase 2 
may have different physics from the better-studied nematic phase 1. 
Nematic phase 1 
 is associated with the field-angle-tuned QCEP of $H_1$. $H_2$ at ambient pressure may be close to its QCEP, since at ambient pressure $\chi''$ shows weak first order behaviour, while susceptibility measurement at 0.59 kbar and higher pressures did not observe any $\chi''$ signal at $H_2$. Regardless of this, a key point is that nematic phase 2 
  does not screen a metamagnetic QCEP, because it is located {\em beside} $H_2$ (see Fig.\  \ref{fig:intro}, \ref{fig:rhovsH}, \ref{fig:1kbar} and \ref{fig:100mK}). Moreover, our finding that the nematic bump is relatively unaffected by pressure also seems to rule out fine tuning to a QCEP as being necessary for formation of this nematic phase.

An obvious explanation for the robustness of nematic phase 2, 
  in terms of a symmetry-breaking Fermi surface reconstruction, 
  would be that the in-plane applied magnetic field already breaks four-fold symmetry 
  via coupling of the electron momentum to the applied field\cite{Puetter2007} 
  as well as through 
  magnetoelastic coupling, which normally stretches a crystal parallel to $\vec{H}$
  and shrinks it perpendicular to $\vec{H}$. 
The resulting distortion of the energy bands would lower the degeneracy of 
  any van Hove singularity in the Brillouin zone from one set of four-fold degenerate,  
  to a pair of two-fold degenerate van Hove singularities, increasing the 
  tendency for the Fermi surface to reconstruct in a two-stage process. The double peaks in $\chi'(H)$ at $H_2$ at low temperatures may arise from such a splitting of the van Hove singularity. However, it is quite clear from the narrowness of the $H_2$ metamagnetic transition compared with the width of nematic phase 2 
  region, that any such splitting is tiny compared with the width nematic phase 2, 
  and that the van Hove singularity remains below the lower boundary of 
 nematic phase 2. 
The physics of nematic phase 2 
remains a mystery. 
There is evidence that the resistivity in nematic phase 2 is affected by the angle between the current and the principal axis. More extended experiments on the magnetoresistance with different angles between the current and the principal axis are needed in order to explore this behaviour. Moreover, quantum oscillation measurements under pressure would help to show how the Fermi surface changes across this nematic phase, which would should also enhance understanding of the underlying physics. 

\section{Summary}
In conclusion, the relation of nematicity and quantum criticality has been studied in Sr$_3$Ru$_2$O$_7$ by applying hydrostatic pressure when the magnetic field is in the \textit{ab}-plane. There is no evidence of a nematic phase at $H_1$ when there is a QCEP induced by hydrostatic pressure. This is in contrast with the appearance of the nematic when the QCEP is obtained by field-angle tuning.
 Another nematic phase, persistent with pressure, does not occur at a metamagnetic quantum critical point. These two findings suggest that, in contrast to quantum critical superconductors, the nematic phase is not driven by quantum criticality. 

\begin{acknowledgements}
We are grateful to H. Y. Kee for helpful discussions. This research was supported by the NSERC (Canada) and EPSRC (UK). SAG was partially supported by the Royal Society (UK), CONICET and ANPCyT (Argentina), and APM holds a Royal Society - Wolfson Research Merit Award.

\end{acknowledgements}

\appendix 
\section{}
In our previous pressure experiment,\cite{Wu2011} \textit{ac} magnetic susceptibility $\chi_{ac}(H,\omega)$ was measured at low frequency.  There are different possible regimes for $\chi_{ac}$ in the low frequency limit depending on the value of $\omega$ compared with $\tau^{-1}$, the inverse characteristic time of the system: if $\omega \ll \tau^{-1}$, $\chi_{ac}$ tends to the isothermal susceptibility, $\chi_T=\left(\frac{\partial M}{\partial H}\right)_T$. On the opposite limit, $\omega \gg \tau^{-1}$, the system has no time to exchange energy with its surroundings and what is measured is the adiabatic susceptibility $\chi_S$, which is usually smaller in size. A good rule of thumb, based on the analysis by Casimir and du Pr\'e,\cite{casimir1938note} is to work in the very low frequency regime where the imaginary part of $\chi_{ac}(H,\omega)$ is negligible.  In this regime, $\chi_{ac}(H,\omega) \approx \left(\frac{\partial M}{\partial H}\right)_T $ and the magnetization $M(H,T)$ can, in principle, be  obtained by simple integration over the field $H$.  In practice, in addition to keeping a low frequency, this is difficult to do because the filling factor of the pick-up coil, and the overall gain of the system are not known with sufficient accuracy. The integrated susceptibility in Fig.\  \ref{fig:chi} of our paper was therefore calculated by the following equation:
\begin{equation}
M(H)=\int (a\chi'_{ac}(H)+b)\, \mathrm{d}H,
\end{equation} 
where $a$ and $b$ were chosen so that $M(H)$ matches  magnetization measurements by Perry et al.\ at 70 mK.\cite{Perry2005}
\begin{figure}[htb]
\includegraphics[width=7.8cm]{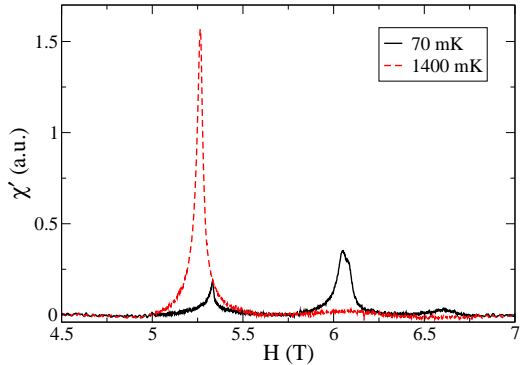} 
\caption{Real part of susceptibility $\chi(H)$ at different temperatures. The second and third peaks in 70 mK data mark the entrance and exit of  nematic phase 2,  
 respectively, with the second peak being weakly split. 
} 
\label{fig:Appchi}
\end{figure}

In addition to the constants $a$ and $b$, there is another adjustment to be made since the rise in $M(H)$ at $H_1$ is of first-order. It can be seen that the first peak is much larger near the critical temperature of 1550 mK (see Fig.\  \ref{fig:Appchi}). At lower temperatures the dynamical response is affected by the physics of a first-order metamagnetic transition, in particular by domain wall pinning, with the consequent growth of the characteristic time, $\tau$.   In this region, the imaginary part of $\chi_{ac}$ is no longer negligible, and the real part of $\chi_{ac}$ decreases towards $\chi_S$. Thus at the lowest temperature, 70 mK, while the size of the second peak is appreciable, the first peak has become very small. In order to compensate for the signal loss and achieve an agreement with the measured $M(H)$, we add a first-order jump at $H_1$ in Fig.\  \ref{fig:chi}.   

\bibliographystyle{phaip}
\pagestyle{plain}
\bibliography{327collection}
\end{document}